\documentclass[prl,twocolumn,showpacs]{revtex4}

\usepackage{amssymb}

\usepackage{graphicx}
\usepackage{amsmath}

\renewcommand\bigskip{}

\begin{document}

\title{Theory of Diffusion Controlled Growth}
\author{R. C. Ball}
\affiliation{Department of Physics, University of Warwick,
        Coventry CV4 7AL, U.K.}

\author{E. Somfai}
\affiliation{Department of Physics, University of Warwick,
        Coventry CV4 7AL, U.K.}

\begin{abstract}
We present a new theoretical framework for Diffusion Limited Aggregation and
associated Dielectric Breakdown Models in two dimensions. Key steps are
understanding how these models interrelate when the ultra-violet cut-off
strategy is changed, the analogy with turbulence and the use of
logarithmic field variables.
Within the simplest, Gaussian, truncation of mode-mode coupling, all
properties can be calculated. The agreement with prior knowledge from
simulations is encouraging, and a new superuniversality of the tip scaling
exponent is both predicted and confirmed.
\end{abstract}


\maketitle

Diffusion Limited Aggregation (DLA) has accumulated an enormous literature
since Witten and Sander first introduced their simulation model of a rigid
cluster growing by the accretion of dilute diffusing particles
\cite{wittensander}. The importance of the model is that it encompasses a
range of problems where growth or interfacial advance is governed by a
conserved gradient flux, that is the local interfacial velocity is given by 
\begin{equation}
v_{n}\propto \left| \partial _{n}\phi \right| ^{\eta },\qquad \nabla
^{2}\phi =0,\text{\qquad }\phi _{\text{interface}}\approx 0  \label{growth}
\end{equation}
where for DLA $\eta =1$~\cite{cargese}. The generalisation to a range of
positive $\eta $ was introduced by Niemeyer, Pietronero and Wiesmann
\cite{DBM} to model dielectric breakdown patterns, but in this letter we
exploit it to support proposed equivalences between models with significantly
different ultraviolet cut-off mechanism. Theoretical interest has been fuelled
by the fractal and multifractal \cite{singularities,hmp} scaling properties of
the clusters produced, with controversial claims
\cite{anomalousscaling,coniglio89,mandelbrot} (and counter-claims
\cite{counterclaims,somfai99,noisereduction}) of anomalous scaling, and by the
longstanding absence here resolved of an overall theortical framework to
understand the problem.

The presence of a cut-off lengthscale $a$ below which the physics dictates
smooth growth is a crucial ingredient of DLA; it is known that otherwise
infinitely sharp cusps develop in the interface within finite time
\cite{bensimonshraiman}. In DLA this cutoff is fixed and set by the size of
accreting particles, but there are other problems where it is set in a more
subtle dynamical way by the surface boundary conditions on the diffusion
field. In dendritic soldification this comes about through competition between
surface energy and diffusion kinetics (with $\eta =1$), leading to 
\begin{equation}
a\propto \left| \partial _{n}\phi \right| ^{-m}  \label{tip}
\end{equation}
with $m=1/2$ at least for those tips not in retreat \cite{langer}. In terms of
$m$, simple DLA corresponds to $m=0$, and in the theory below in two
dimensions we will map onto the case where $a$ is such that each growing tip
has fixed integrated flux, corresponding to $m=1$.

It is central to fractal (and multifractal) behaviour in DLA that the measure
given by the diffusion flux [density] $\partial _{n}\phi $ onto the interface
has singularities \cite{hmp,singularities}, such that the integrated flux onto
the growth within distance $r$ of a singular point is given by 
\begin{equation}
\mu (r)\sim r^{\alpha }.  \label{scaling}
\end{equation}
Applying this phenomenology to the scaling around growing tips, we can
establish an equivalence between models at different $\eta $ and $m$ by
requiring that \textit{the relative advance rates of different growing tips
are matched}. Consider two competing tips labelled $1,2$, for two growths with
the same overall geometry but growing governed by parameters $(\eta ,m)$ and
$(\eta ^{\prime },m^{\prime })$ respectively. For tip 1 we will have tip
radius $a_{1}$ and flux density $j_{1}$ which are matched between the two
different models by $j_{1}^{\prime }a_{1}^{\prime d-1}/a_{1}^{\prime \alpha
}=j_{1}a_{1}^{d-1}/a_{1}^{\alpha }$ and similarly for tip 2, whilst the two
tips are interrelated by $a_{1}^{\prime }j_{1}^{\prime m^{\prime
}}=a_{2}^{\prime }j_{2}^{\prime m^{\prime }}$ and similarly for the unprimed
quantities. If we insist that their advance velocities are in the same ratio
in both models this requires $(j_{1}/j_{2})^{\eta }=(j_{1}^{\prime
}/j_{2}^{\prime })^{\eta ^{\prime }}$, which forces the parameter relation 
\begin{equation}
\frac{1+m(1+\alpha -d)}{\eta }=\frac{1+m^{\prime }(1+\alpha -d)}{\eta
^{\prime }}.  \label{equivalence}
\end{equation}
For the two models to be equivalent in the relative velocities of all tips
requires their parameters be related as above, where $\alpha $ is the
singularity exponent associated with growing tips which we take to be the same
as we are matching the geometry at scales above the cut-offs.

Although we have not strictly proved the equivalence of the models related
above, we have shown that any such relationship must follow
Eq.~(\ref{equivalence}) and we will assume in the rest of this letter that
this equivalence holds. All such models are then classifiable in terms of a
convenient reference such as $\eta _{0}$, the equivalent $\eta $ when $m=0$,
corresponding to the original Dielectric Breakdown Model. For example
dendritic solidification with $\eta =1$ and $m=1/2$ corresponds to $\eta
_{0}=\frac{2}{3+\alpha -d}$: it is thus not equivalent to DLA, but to another
member of the DBM class. Another puzzle resolved by our classification is a
recent study showing conflicting scaling between DLA and different limits of a
'laplacian growth' model \cite{laplaciangrowthmodel}.  In the present
terminology the latter model corresponds to $m=-1$ and its two limits of low
and high coverage of the growing surface per growth step have $\eta =3$ and
$\eta =1$ respectively. Using $\alpha =0.7$ (see below) these map through
Eq.~(\ref{equivalence}) into $\eta _{0}=2.31$ and $\eta _{0}=0.77$
respectively, so the way their scaling brackets that of DLA is quite expected.

\begin{figure}
\resizebox{0.9\columnwidth}{!}{\includegraphics*{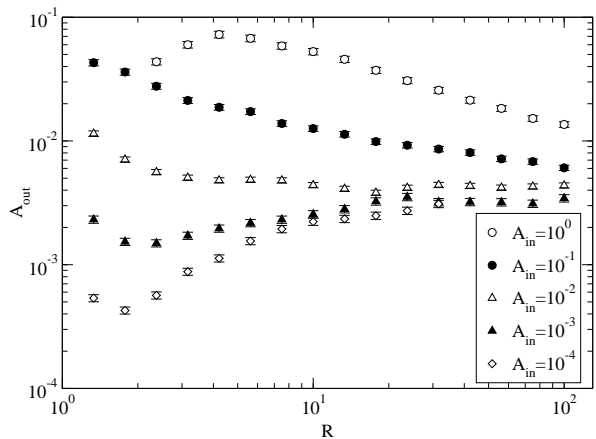}}
\caption{\label{fig:noise}
The size fluctuations (``output noise'') $A_{\text{out}} = (\delta N/N)^2$,
measured at fixed radius $R$ for DLA clusters grown with off-lattice noise
reduction \cite{noisereduction} at various (``input'') noise levels
$A_{\text{in}}$. For low  $A_{\text{in}}$, $A_{\text{out}}$ self organizes
from below in the manner of a turbulent system.
}
\end{figure}

DLA and DBM have hitherto been regarded as models in statistical physics, in
that the local advance rate in Eq.~(\ref{growth}) has been implemented as the
probabilty per unit time for the growth locally to make some unit of advance,
entailing an inherent shot noise. Here we argue that diffusion controlled
growth is a problem of turbulence type, with noise self-organising from
minimal input.  The data in Fig.~\ref{fig:noise} show how the relative
fluctuations can approach their limiting value from below as well as from
above.

The new ideas above, that we can balance changing the cut-off exponent $m$ by
adjustment of $\eta$, and that noise can be left to self organise, \ are the
key to a new theoretical formulation of the problem, at least in two
dimensions of space to which we now specialise. In two dimensions the Laplace
equation in (\ref{growth}) can be solved in terms of a conformal
transformation between the physical plane of $z=x+iy$ and the plane of complex
potential $\omega =\phi +i\theta $ in which we take the growing interface to
be mapped into the periodic interval $\theta =[0,2\pi ),\phi =0$ and the
region outside of the growth mapped onto $\phi >0$. Then adapting reference
\cite{bensimonshraiman}, we have for the dynamics of the interface following
Eq.~(\ref{growth}), 
\begin{equation}
\frac{\partial z(\theta )}{\partial t}=-i\frac{\partial z}{\partial \theta }
\mathcal{P}\bigg[\left| \frac{\partial \theta }{\partial z}\right| ^{1+\eta }
\bigg].
\label{bseqn}
\end{equation}
The linear operator $\mathcal{P}$ is most simply described in terms of Fourier
transforms: $\mathcal{P}[\sum_{k}e^{-ik\theta }f_{k}]=\sum_{k}P[k]e^{-ik\theta
}f_{k}=f_{0}+2\sum_{k=1}^{K}e^{-ik\theta }f_{k}$, where we have introduced
here an upper cut-off wavevector $K$. It is easily shown that on scales of
$\theta$ greater than $K^{-1}$ a smooth interface is linearly unstable with
respect to corrugation for $\eta >0$ (the Mullins-Sekerka instability
\cite{mullinssekerka}), whereas for scales of $\theta $ less than $K^{-1}$ the
equation drives smooth behaviour (corresponding locally to the case $\eta
=-1$). This cutoff on a scale of $\theta $, the cumulative integral of flux,
corresponds in terms of tip radii and flux densities to $aj\thickapprox
K^{-1}$, that is an $m=1$ cutoff law. Thus the parameter $\eta $ in
Eq.~(\ref{bseqn}) is more specifically $\eta _{1}=\alpha \eta _{0}$, using
Eq.~(\ref{equivalence}) with $d=2$.

\begin{figure}
\resizebox{0.9\columnwidth}{!}{\includegraphics*{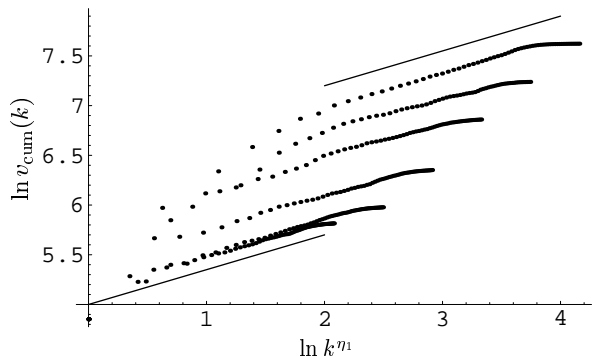}}
\caption{\label{fig:alphadata}
Cumulative contribution to the mean growth velocity plotted against wavevector
as $k^ {\protect\eta _{1}}$with logarithmic scales. The data are (bottom to
top) for $\protect\eta _{1}=0.5,0.6,0.7,0.8,0.9,1.0$ and all exhibit a common
power law slope $1/\protect\alpha -1\approx 0.35\pm 0.04$ per the guidelines
shown.
}
\end{figure}

We have made a numerical test of Eq.~(\ref{bseqn}) and the equivalence
(\ref{equivalence}), with disorder supplied only through the initial
condition, by applying them to the case of growth along a channel with
periodic boundary conditions (cylinder). For this case analyticity of the
conformal map requires that $z(\theta )=i\theta +\sum_{k\geq
0}z_{k}e^{-ik\theta }$ and the overall advance rate of the growth reduces to
$\frac{\partial z_{0}}{\partial t}=\left( \left| \frac{\partial \theta }
{\partial z}\right| ^{1+\eta _{1}}\right) _{0}$, which we can compare to the
expected scaling of tip velocity with the cutoff, $v\thicksim K^{(1-\alpha
)\eta _{1}/\alpha }$. It is convenient to change variables to $\psi =(\partial
z/\partial \theta)^{-(1+\eta _{1})/2}$, in terms of which we obtain 
\begin{equation}
\frac{\partial \psi }{\partial t}=-i\frac{\partial \psi }{\partial \theta }
\mathcal{P}[\psi \overline{\psi }]+iy\psi \frac{\partial }{\partial \theta }
\mathcal{P}[\psi \overline{\psi }]  \label{bseqncubic}
\end{equation}
where $y=(1+\eta _{1})/2$ and the tri-linear form of the RHS enables us to
compute numerically the motion within a purely Fourier representation.
Figure~\ref{fig:alphadata} shows the measured variation of $\sum_{j<k}\left|
\psi _{j}\right| ^{2}$ vs $k^{\eta _{1}}$: this is expected to exhibit a power
law with exponent $(1/\alpha -1)$ and the observed slope plotted in this way
is surprisingly independent of $\eta _{1}$.

The most important result of our numerical study of Eq.~(\ref{bseqncubic}) is
that this clearly does self-organise into statistical scaling behaviour, given
disorder from only the initial conditions.  However the numerical results are
also remarkable, as we obtain $\alpha \approx 0.74\pm 0.02$ with no
significant dependence on $\eta_{1}$ in the range studied. This not only
agrees reasonably with the value $\alpha =D-1=0.71$ known from large direct
simulations of DLA \cite{DLAdimension,channelxsi}, but also appears to imply a
deeper unversality which we will see is replicated in our analytic theory
below.

We now turn to a theoretical analysis of Eq.~(\ref{bseqn}), for which a
primary requirement is that we must obtain results explicitly independent of
the cut-off as $K\rightarrow \infty $. This is hard because we have already
seen that the mean advance rate of the interface diverges as a power of $K$,
and on fractal scaling grounds one would expect the same divergent factor to
appear in the rate of change of simple variables such as $z_{k}$ or $\psi
_{k}$. One can of course take ratios of rates of change and look to order
terms such that divergences cancel, but to make this work we have been forced
to introduce yet another change of variables, 
\begin{equation}
-i\frac{\partial z}{\partial \theta }=\exp \left( -\lambda (\theta )\right)
=\exp \left( -\sum_{k>0}\lambda _{k}e^{-ik\theta }\right) ,
\label{logvars}
\end{equation}
which corresponds to Fourier decomposing the logarithm of the flux density.
The key to the success of these variables is that they decompose the flux
density itself multiplicatively and, as we shall see, quite naturally capture
its multifractal behaviour. In terms of these 'logarithmic variables' the
equation of motion becomes 
\begin{equation}
\frac{\partial \lambda _{k}}{\partial t}=-\sum_{j\leq k}(k-j)\lambda
_{k-j}P(j)\left( e^{y(\lambda +\overline{\lambda })}\right)_{j}+2k\left(
e^{y(\lambda +\overline{\lambda })}\right)_{k}  \label{logeom}
\end{equation}
where subscripts on bracketed expressions imply the taking of a Fourier
component, by analogy with $\lambda_{k}$. The advance rate of the mean
interface is given in these variables by $\frac{ \partial z_{0}}{\partial
t}=\left( e^{y(\lambda +\overline{\lambda })}\right) _{0}$.

Now let us suppose some ignorance of the initial conditions and describe the
system in terms of a joint probability distribution over the $\lambda _{k}$,
and let us denote averages over this [unknown] distribution by $\left\langle
..\right\rangle $. We can in principle determine the distribution through its
moments, whose evolution we now compute. For simplicity we assume
translational invariance with respect to $\theta $, so that only moments of
zero total wavevector need be considered, of which the lowest gives: 
$
\frac{\partial }{\partial t}\left\langle \lambda _{k}\overline{\lambda }%
_{k}\right\rangle =\left( -\sum\limits_{j\leq k}(k-j)P(j)\left\langle
\lambda _{k-j}\overline{\lambda }_{k}e_{j}^{y(\lambda +\overline{\lambda }%
)}\right\rangle +2k\left\langle \overline{\lambda }_{k}e_{k}^{y(\lambda +%
\overline{\lambda })}\right\rangle \right) +\left( \text{c. conj.}\right) 
$. 
All of the higher moments lead to the same form of averages on the RHS, 
$\left\langle \text{multinomial(}\lambda ,\overline{\lambda }\text{)}
e^{y(\lambda +\overline{\lambda })}\right\rangle $, and all of these terms
are conveniently expressed in terms of cumulants \cite{cumulants}, using
the identities $\left\langle Xe^{W}\right\rangle /\left\langle
e^{W}\right\rangle =\left\langle Xe^{W}\right\rangle _{c}$,
$\left\langle XYe^{W}\right\rangle /\left\langle e^{W}\right\rangle
=\left\langle XYe^{W}\right\rangle _{c}+\left\langle Xe^{W}\right\rangle
_{c}\left\langle Ye^{W}\right\rangle _{c}$, etc. 
The key helpful feature is that the
expressions we require all naturally divide by one factor of $\left\langle
e^{y(\lambda +\overline{\lambda })}\right\rangle = \frac{\partial }{
\partial t}\left\langle z_{0}\right\rangle $, which is what we
sought in order to remove divergences.

To obtain tractable results we need to introduce some closure approximation(s)
and we present here the simplest, neglecting all cumulants higher than the
second, equivalent to assuming a joint Gaussian distribution (of zero mean)
for $\lambda $. This is entirely characterised by its second moments
$S(k)=\left\langle \lambda _{k}\overline{ \lambda }_{k}\right\rangle $ which
by Eq.~(\ref{logeom}) we find evolve according to $\partial S(k)/\partial
\left\langle z_{0}\right\rangle
=-kS(k)-y^{2}kS(k)^{2}-2y^{2}\sum_{j<k}jS(j)S(k)+2ykS(k)$. This in turn
approaches a stable steady state solution where 
\begin{equation}
S(k)=\frac{2y-1}{y^{2}}k^{-1}\,,\quad k\text{ odd;} \quad\,\,
S(k)=0\,,\quad k\text{ even.}  \label{variances}
\end{equation}
The absence of even $k$ is readily interpreted in terms of the dominance of
one major finger and one major fjord.

\begin{figure}
\resizebox{0.9\columnwidth}{!}{\includegraphics*{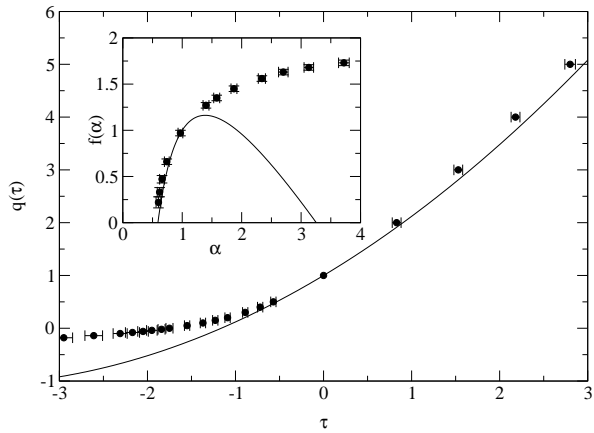}}
\caption{\label{fig:tauq}
Multifractal spectra from the Gaussian theory ($\alpha=2/3$), compared to
measured values for DLA \cite{f(a)data}. Agreement is excellent for the active
region $\tau\geq 0$, $\alpha\leq 1$, and there are no adjustable parameters.
}
\end{figure}

Within the Gaussian approximation and its predicted variances
(\ref{variances}) we can now compute all [static] properties of diffusion
controlled growth, in a channel and (see later discussion) also in radial
geometry. The multifractal spectrum of the harmonic measure follows from
computing the general moment \cite{hmp} $\left\langle \left| \frac{\partial
\theta }{\partial z}\right| ^{-\tau }\right\rangle =\left\langle e^{(\lambda
+\bar{\lambda})\tau /2}\right\rangle =\exp \left( \tau
^{2}/4\sum_{k}^{K}S(k)\right) \simeq K^{q(\tau )-1-\tau }$, leading to 
\begin{equation}
q(\tau )=1+\tau +\tau ^{2}\frac{\eta _{1}}{2\left( 1+\eta _{1}\right) ^{2}}
\label{moments}
\end{equation}
and it is easy to see that any closure scheme based on keeping cumulants of
$\lambda $ up to some finite order leads to a polynomial truncation of $q(\tau
)$. From the Legendre Transform of the inverse function $\tau (q)$ we obtain
the corresponding spectrum of singularities, 
\begin{equation}
f(\alpha )=2-\frac{1}{\alpha}+\frac{1}{2}\left( \eta _{1}+\frac{1}{\eta _{1}}
\right) \left( 2-\alpha -\frac{1}{\alpha }\right)  \label{falpha}
\end{equation}
which in Fig.~\ref{fig:tauq} is compared to measured data for DLA
\cite{f(a)data}, which later measurements \cite{jensen02} reinforce.  For the
region of active growth $\alpha\leq 1$ ($q\geq 0$) the theory is
quantitatively accurate. At $\alpha=1$ it conforms to Makarov's theorem
\cite{makarov}, and in contrast to the Screened Growth Model
\cite{screenedgrowth} it does this without adjustment. For $\alpha>1$ the
spectrum only qualitatively the right shape, and for such screened regions our
equations based on tip scaling may not hold.

Although the multifractal moments depend significantly on the input parameter
$\eta _{1}$, the tip scaling exponent $\alpha $ turns out to be independent of
this and in close agreement with our numerical results.  Matching the expected
scaling of the mean velocity (as used to measure $\alpha$ above) to that of
the multifractal moment with $\tau =-(1+\eta_{1})$ leads direcly to $\alpha
=2/3$ independent of $\eta _{1}$.  This is a remarkable success for the
Gaussian Theory to predict this hitherto unexpected result so closely.

The multifractal spectrum suggests that the Gaussian approximation is good in
the growth zone, so we have computed the penetration depth as a further test.
For growth in the channel we define relative penetration depth $\Xi $ as the
standard deviation of depth $\Re(z)$ along the chanel, computed over the
harmonic measure, divided by the width of the channel. This leads to $\left(
2\pi \Xi \right) ^{2}=\left\langle z\overline{z} \right\rangle
/2=\sum_{k>0}k^{-2}\left\langle \left| e_{k}^{\lambda }\right|
^{2}\right\rangle /2$ where the required averages can all be computed in the
approximation of Gaussian distributed $\lambda $. Using $\eta _{1}=2/3$
corresponding to DLA this leads to $\Xi _{\text{theory}}=0.13$, \ compared to
$\Xi _{\text{DLA}}=0.14$ from direct simulations of DLA growth in a periodic
channel \cite{channelxsi}.

All of the new theory is readily extended to growth from a point seed in
radial geometry. The multifractal spectrum turns out to be unchanged, in
accordance with expectations from universality.  The penetration depth
relative to radius gives $\Xi _{\text{theory}}=0.20$ for radial DLA, compared
to our recently published extropolation from simulations, $\Xi
_{\text{DLA}}=0.13$ \cite{somfai99}.

For DLA and its associated Dielectric Breakdown Models we have shown a
theoretical framework which is complete in the sense that essentially all
measurable quantities can be calculated. For amplitude factors such as the
relative penetration depth there is no theoretical precedent. For the full
spectrum of exponents the advance over the Screened Growth Model is the
elimination of fitting parameters. For the exponent $\alpha _{ \text{tip}}$ we
have in the Gaussian approximation a striking new result that this is
independent of $\eta $, which begs direct confirmation by (expensive)
particle-based simulations. However for DLA in particular we have not yet
improved on the best theoretical value of $\alpha _{\text{tip}} $, which
remains $1/\sqrt{2}\approx 0.71$ from the Cone Angle Approximation
\cite{coneangle}.

Within DLA and DBM we look forward to calculating more properties such as the
response to anisotropy, which is fairly readily incorporated into our
equations of motion. A more challenging avenue is to improve on the Gaussian
approximation which we have used to obtain explicit theoretical results.
Truncating at a cumulant of higher order than the second is hard, and more
seriously it does not correspond to a positive (semi-)definite probability
distribution. An alternative route of improvement which we are exploring is
closure at the level of the full multifractal spectrum.

There are possibilities for wider application of ideas in this letter, where
we have formulated DLA and DBM as a turbulent dynamics governed by a complex
scalar field in 1+1 dimensions. Decomposing this field multiplicatively
(through Fourier representation of its logarithm) was the crucial step to
obtain renormalisable equations and theoretical access to the multifractal
behaviour, even though other representations offered equations of motion
(\ref{bseqncubic}) with weaker non-linearity. It is natural to speculate
whether the same strategy might apply to turbulent problems more widely, where
the key issue appears to be identifying suitable fields to decompose
multiplicatively which are of local physical significance, and subject to
closed equations of motion.

\begin{acknowledgments}
This research has been supported by a Marie Curie Fellowship of the EC
programme ``Improving Human Potential'' under contract number
HPMF-CT-2000-00800.
\end{acknowledgments}

\end{document}